\documentstyle[aps,preprint]{revtex}
\begin{document}

\draft
\tightenlines
\makebox[\textwidth][r]{SNUTP 97-146}
\begin{center}
{\large\bf
Chiral symmetry breaking by a non-Abelian external field in
(2+1)-dimensions
} 
\vskip 0.5in
V. P. Gusynin$^{1}$\footnote{Email 
address:$\>$\tt vgusynin@gluk.apc.org},  
D. K. Hong$^{2}$\footnote{Email address:
$\>$\tt dkhong@hyowon.cc.pusan.ac.kr},
and I. A. Shovkovy$^{1,3}$\footnote{Email address:
$\>$\tt igor@physics.uc.edu} 
\\
\vskip 0.2in 
\it 
$^{1}$Bogolyubov Institute for Theoretical Physics
\protect\\
252143 Kiev, Ukraine\protect\\
$^{2}$Department of Physics, Pusan National
University \protect\\
Pusan 609-735, Korea\protect\\
$^{3}$Physics Department, University of Cincinnati
\protect\\  OH 45221-0011, USA\protect\\
\end{center}


\begin{abstract} 
We investigate the effect of a constant external non-Abelian 
field on chiral symmetry breaking in a (2+1)-dimensional 
Nambu-Jona-Lasinio model and in 3D QCD by solving the gap 
equation and the Bethe-Salpeter equation, and also by RG 
analysis.  In the (2+1)-dimensional NJL model chiral symmetry 
breaking occurs for any weak coupling constant but in 3D QCD
catalysis of chiral symmetry breaking does not occur. 
\end{abstract}


Though there are increasing evidences for chiral symmetry breaking
($\chi$SB) in QCD~\cite{qcd}, we still do not understand clearly
how it occurs, due to a lack of comprehensive understanding the
QCD vacuum.  One of the popular analytic methods to study $\chi$SB
in QCD is by solving the Schwinger-Dyson (SD) equation for the
Green functions, from whose solution one can extract the structure
of the vacuum~\cite{sd}.  Most attention in SD analysis has been
focused so far on the effect of the strong coupling in QCD; if the
coupling is bigger than a critical coupling, the SD equation admits
stable $\chi$SB solutions.  But, the effect of self-interacting
gluons in SD analysis  has not been seriously considered except in
the running coupling, even though it has been conjectured that the
monopole condensate in QCD leads to color confinement and thus
$\chi$SB~\cite{thooft}, which is now shown rigorously in $N=2$
supersymmetric QCD~\cite{seiberg}.

Since the quark condensate under an external gluon field
$A_{\mu}^a$ is related to the density of eigenvalue of Dirac
operator $i\mathord{\not\mathrel{D(A)}}$ at zero
eigenvalue~\cite{banks},
\begin{equation}
 \left<0\right|\bar qq\left|0\right>^A=-\pi\rho_A(0),
\end{equation}
it is conceivable that a certain gluon field configuration which
gives a constant eigenvalue density at small eigenvalue like a free
(0+1)-dimensional spectrum is a source of $\chi$SB.  On the other
hand, it has been recently found  that an external magnetic field
catalyzes chiral symmetry breaking in the Nambu-Jona-Lasinio (NJL)
model~\cite{NJL} and in quantum electrodynamics~\cite{gusynin,leung,my}.
The fact that a constant magnetic field enhances fermion dynamical
masses was known from the papers~\cite{KKK}. Studying (2+1)-dimensional
NJL model with discrete chiral symmetry, it was shown \cite{Klim} also
that chiral symmetry turns out spontaneously broken for any $B\neq 0$.
However, what was shown in \cite{NJL} is not just that the magnetic 
field enhances dynamical fermion masses created by supercritical 
NJL interaction, but that it catalyzes generating the mass even at 
the weakest attractive interaction and this effect is universal, i.e.,
model independent in 3+1 and 2+1 dimensions\footnote{For recent 
application of this effect in condensed matter physics 
see~\cite{mavromatos}}. The physical reason for the magnetic 
catalysis of $\chi$SB is that, in the presence of external magnetic  
field $\vec B=B\hat z$, dimensional reduction occurs in fermion
motion. The fermion spectrum becomes one-dimensional, 
$E=\pm|p_z|$, at energy much lower than the Landau gap 
$E\ll \sqrt{|eB|}$ and, for one-dimensional fermions, an 
arbitrarily weak attraction at scale
$\mu=\sqrt{|eB|}$ leads to a strong attraction in the infrared
region~\cite{my}.  It is important that in \cite{NJL,gusynin}
continuous chiral symmetry was considered, that is far nontrivial
problem since the appearance of Nambu-Goldstone bosons
along with reduction to $D=0+1$ or $D=1+1$ dimensions might
wash out symmetry breaking, as happens, for example, in the chiral
(1+1)-dimensional Gross-Neveu model~\cite{witten}. However, it was
shown that notwithstanding dimensional reduction Goldstone boson
propagators keep their three or four dimensional form.

In studying $\chi$SB in QCD, it is natural to look for a gluon
field that mimics a magnetic field and gives a (1+1)-dimensional
spectrum for the Dirac operator. We consider this in a
(2+1)-dimensional NJL model and in a (2+1)-dimensional $SU(2)$
gauge theory for simplicity. These theories in 2+1 dimensions  have
been studied intensively recently, since they  can serve not only as
theoretical laboratories for investigating  aspects of chiral symmetry
breaking in QCD but also describe the high temperature thermodynamics
of 4-dimensional theories as well as certain planar condensed matter
systems.

A constant and uniform chromomagnetic field for a non-Abelian gauge theory
can be described by two different types of gauge potentials~\cite
{Brown,Leutwyler}, a feature not available for the abelian case.
As is known these two different potentials lead to two different
energy spectra for fermions in such a field~\cite{Brown,Ebert}.
The non-Abelian external field which mimics a magnetic field and has
the largest global symmetry including  explicit translational 
invariance, is described by a constant gauge potential
\begin{equation}
A_{\mu}={1\over2}A_{\mu}^a\tau^a,\quad{\rm with}~
A_1^1=A_2^2=\sqrt{H/g},~{\rm and~ others}=0,
\label{potential}
\end{equation}
where $\tau^a$'s are the Pauli matrices ($a=1,2,3$),
$g$ is the gauge coupling constant, and $H$ is a constant.
Then, the field strength tensor,
\begin{equation}
F_{\mu\nu}^a{\tau^a\over2}={1\over ig}\left[D_{\mu},D_{\nu}\right],
\end{equation}
so that $F_{12}^3=-F_{21}^3=-H$ and others are zero, where
the covariant derivative $D_{\mu}=\partial_{\mu}+igA_{\mu}^a\tau^a/2$.
Under the external field, Eq.~(\ref{potential}), the fermion
propagator satisfies
\begin{equation}
 \left[\gamma^{\mu}\left(i\partial_{\mu}-gA_{\mu}\right)-m
\right]S(x-y)=i\delta(x-y),
\end{equation}
where $m$ is a fermion mass and we take the four-component reducible
(2+1)-dimensional spinor for fermions. To find the fermion propagator
we look for a solution of the form
\begin{equation}
S(x-y)=i\left[\gamma^\mu
\left(i\partial_\mu-gA_\mu\right)+m\right]\Delta(x-y),
\end{equation} 
where the Fourier transform of $\Delta(x-y)$ satisfies 
the following equation:
\begin{equation}
\left[P_\mu P^\mu-{g\over4}\sigma^{\mu\nu}F^a_{\mu\nu}
\tau^a-m^2\right]\Delta(p)=1,
\label{eq:delta}
\end{equation}
where
\begin{equation}
P_\mu=p_\mu-gA_\mu,\qquad \sigma_{\mu\nu}
={i\over2}[\gamma^\mu,\gamma^\nu],\qquad 
\{\gamma^\mu,\gamma^\nu\}=2g^{\mu\nu}.
\end{equation}
With our choice of the vector potential (\ref{potential}) 
the equation (\ref{eq:delta}) simplifies to
\begin{equation}
\left[p^2-2h(\vec p\cdot\vec\tau)-2h^2
+2h^2\tau^3\sigma^{12}-m^2\right]\Delta(p)=1,
\end{equation}
where we introduced the notation $h=\sqrt{gH}/2$. 
The last equation can easily be solved and we find:
\begin{equation}
 \Delta(p)= \frac{1}{Q_1Q_2}
\left[p^2-2h^2+2h\vec{p}\cdot\vec{\tau}
-2h^2\tau^3\sigma^{12}-m^2\right],
\end{equation}
where
\begin{eqnarray}
Q_{1,2}(p)&=&p_0^2-m^2-\left[ \sqrt{\vec{p}^2+h^2}\pm h \right]^2.
\label{2+1Q's}
\end{eqnarray}
Finally, the Fourier transform of the fermion propagator reads
\begin{equation}
S(p)=i\left[\gamma^\mu\hat P_\mu+m\right]\Delta(p).
\end{equation}
We see that if $m=0$ the pole of the propagator occurs at
\begin{equation}
 p_0=\pm(\sqrt{{\vec p}^2+h^2}\pm h).
\end{equation}
At low momenta, $|\vec p|\ll h$, the spectrum changes drastically, 
it has a non-relativistic branch: $p_0=\pm{\vec p}^2/(2h)$. It occurs 
that just this mode is responsible for chiral symmetry breaking 
in NJL model.

Now, we consider the (2+1)-dimensional NJL model with $N$
four-component fermions in the presence of the above non-Abelian
field, described by the Lagrangian density
\begin{equation}
 {\cal L}=\bar\psi i\gamma^{\mu}D_{\mu}\psi
+{G\over2N}\left[\left(\bar\psi\psi\right)^2+
\left(\bar\psi i\gamma_5\psi\right)^2
+\left(\bar\psi \gamma_3\psi\right)^2\right]
\end{equation}
which has $U(2N)$ symmetry. In terms of auxiliary fields $\sigma$,
$\tau$ and $\pi$ it can be rewritten as
\begin{equation}
{\cal L}^{\prime}=\bar\psi i\gamma^{\mu}D_{\mu}\psi
-\bar\psi\left(\sigma+\gamma^3\tau+i\gamma^5\pi\right)\psi
-{N\over2G}\left(\sigma^2+\pi^2+\tau^2\right).
\label{lagrangian}
\end{equation}
In the large $N$ limit the vacuum is determined by the 
stationary point of the effective action, obtained by integrating
over the fermions:
\begin{equation}
 e^{iS_{\rm eff}(\sigma,\tau,\pi)}=\int\left[d\psi d\bar\psi\right]
e^{i\int d^3x {\cal L}^{\prime}},
\end{equation}
or
\begin{equation}
S_{\rm eff}=-\frac{N}{2G}\int d^3x(\sigma^2+\pi^2+\tau^2)-iTr\log
[i\gamma^\mu D_\mu-(\sigma+i\gamma_5\pi+\gamma_3\tau)].
\end{equation}
Since the vacuum should respect translational invariance, we
need to calculate the effective action for constant auxiliary
fields to find the vacuum. In this case, the effective action is
just $S_{\rm eff}(\sigma,\tau,\pi)=-V(\sigma,\tau,\pi)L^2T$, where
$L^2T$ is the spacetime volume and $V$ is the effective potential.
The potential $V$ depends only on the chiral invariant $\rho^2=\sigma^2+
\pi^2+\tau^2$ and for its calculation it is sufficient to consider
a configuration with $\tau=\pi=0$ and $\sigma=const$. In the
proper-time formalism we get for $V(\sigma)$:
\begin{equation}
V(\sigma)=\frac{N\sigma^2}{2G}-{i\over2}\int\limits_0^\infty\frac
{ds}{s}\int\frac{d^3p}{(2\pi)^3}tr e^{is[P_\mu P^\mu-{g\over4}
\sigma^{\mu\nu}F^a_{\mu\nu}\tau^a-\sigma^2]}.
\label{effpoten}
\end{equation}
 
The matrix in the exponent of (\ref{effpoten}) has two eigenvalues 
(compare with (\ref{2+1Q's}))
\begin{equation}
\lambda_{1,2}=p_0^2-\sigma^2-[\sqrt{{\vec p}^2+h^2}\pm h]^2,
\end{equation}
each of which has $4N$ degeneracy. Thus we write
\begin{eqnarray}
V(\sigma)&=&\frac{N\sigma^2}{2G}+2N\int\limits_{1/\Lambda^2}^\infty
\frac{ds}{s}\int\frac{d^3p}{(2\pi)^3}\left[
e^{-s[p_3^2+\sigma^2+(\sqrt{{\vec p}^2+h^2}-h)^2]}\right.\nonumber\\
&+&\left. e^{-s[p_3^2+\sigma^2+(\sqrt{{\vec p}^2+h^2}+h)^2]}
-2e^{-s(p_3^2+{\vec p}^2)}\right],
\end{eqnarray}
where we made a rotation to Euclidean space, subtracted the part
corresponding to free fermions $(\sigma=0,h=0)$ and introduced the
ultraviolet cutoff $\Lambda$.
The effective potential up to the order of $\Lambda^0$ is computed
to be
\begin{equation}
 V(\rho)={N\rho^2\over 2G}+{N\over\pi}\left[
-{\rho^2\Lambda\over\sqrt{\pi}}+{\rho^3+
\left(\rho^2+4h^2\right)^{3/2}\over3}-h^2\sqrt{\rho^2+4h^2}
-{\rho^2h\over2}\ln\left({2h+\sqrt{\rho^2+4h^2}\over\rho}\right)
\right],
\label{finpoten}
\end{equation}
where we substituted $\sigma$ by the chiral invariant $\rho$.
The expression (\ref{finpoten}) agrees with that one
derived in~\cite{nonabnjl}, except the $SU(3)$ color group was
considered there. The stationary condition for the effective 
potential gives the gap equation: 
\begin{equation}
 0={N\rho\over G}+{N\over\pi}\left[-{2\rho\Lambda\over\sqrt{\pi}}
+\rho^2+\rho\left(\rho^2+4h^2\right)^{1/2}
-\rho h \ln\left({2h+\sqrt{\rho^2+4h^2}\over\rho}\right)\right].
\label{gapeq}
\end{equation}
As $h\to 0$, we recover the known gap equation
\begin{equation}
\rho^2=\rho\Lambda\left({1\over\sqrt\pi}-\frac{\pi}{2\Lambda G}\right).
\label{knowngap}
\end{equation}
It admits a nontrivial solution only if the coupling $G$ is 
supercritical, $G>G_c=\pi^{3/2}/2\Lambda$. The magnetic field 
changes the situation: at $h\neq 0$, it is easy to see that a 
nontrivial solution exists at all $G>0$. The reason for this is 
that the magnetic field enhances the interaction in the
infrared region: at $h\neq 0$ the last term in (\ref{gapeq}) is 
responsible for the existence of a solution at any $G>0$.
The nontrivial solution always minimizes the effective potential, 
if the four-Fermi coupling $G>0$, and therefore is the vacuum
solution. Thus, unlike the NJL model with $h=0$, chiral symmetry
breaking occurs for any value of $G>0$. In particular, when
$G\to0$, 
\begin{equation} \bar\rho=4h \exp\left(-{\pi\over hG}\right),
\label{mass}
\end{equation}
while $\bar\rho\simeq 0.4h$ when $G\alt G_c$,
the critical coupling for chiral symmetry breaking 
of the NJL model in the absence of the external field. Since 
$\bar\rho$ defines the fermion mass in
the Lagrangian density~(\ref{lagrangian}), we find 
that the fermions get a dynamical mass, which is 
$m_{\rm dyn}=4h \exp\left[-\pi/(hG)\right]$ for weak coupling.
To analyze the supercritical region $G\agt G_c$, it is useful 
to introduce the dynamical mass $m_0$, defined by the 
Eq.(\ref{knowngap}). For weak fields ($h\ll m_0$), we find
\begin{equation}
m_{\rm dyn}=\bar\rho=m_0\left(1+\frac{3h^2}{4m_0^2}\right),
\end{equation}
i.e. $m_{\rm dyn}$ increases with $h$.

If the cutoff $\Lambda=h$, the calculations get much simpler and
the reason for the catalysis is easy to understand. Let us now
introduce a dimensionless coupling $\lambda\equiv hG$. Then the 
effective potential becomes ($\rho\ll h$)
\begin{equation}
 V(\rho)=Nh\left[\frac{2h^2}{3\pi}+{\rho^2\over
2\lambda}+{1\over4\pi}\rho^2\left(\ln{\rho^2\over 16h^2}
-{4\over\sqrt{\pi}}+3\right)\right]+O(1),
\end{equation}
from which we can read off the scaling behavior of the coupling $G$
by taking a different renormalization  point as
\begin{equation}
 {1\over \lambda(\mu)}\equiv (Nh)^{-1}
\left. {d^2V\over d\rho^2}\right|_{\mu}
={1\over \lambda}+{1\over 2\pi}\left(\ln{\mu^2\over 16h^2}
-{4\over\sqrt{\pi}}+6\right).
\end{equation}
The dynamical mass of a fermion, which will be invariant
under the change of the renormalization point, is of the
order of
\begin{equation}
 m_{\rm dyn}\simeq 4h \exp \left(-{\pi\over hG}+\pi+\frac{2}
{\sqrt\pi}-3\right).
\label{rg}
\end{equation}
This result is very similar to magnetic catalysis, where
an arbitrarily weak  magnetic field catalyzes chiral symmetry
breaking, though the motion of fermions at low energy is not
dynamically reduced to one dimension by the uniform chromomagnetic
field. The reason for catalysis of $\chi$SB by the uniform
chromomagnetic field is that the spectrum of fermions changes at
low energies, $E\ll h$, so that the scaling dimension of fermions is
the same as that of fermions in (1+1)-dimensions and thus the 
four-Fermi interaction becomes a relevant operator. This can be 
easily seen by performing the renormalization group analysis similar to that  
which was applied to the ordinary magnetic catalysis~\cite{my}.

Let us expand the fermion field as
\begin{equation}
\Psi(\vec r, t)=\int_k\sum_A\psi_A(t,k)
e^{i{\vec k}{\vec r}}u_A(\vec k),
\end{equation}
where $u_A(\vec k)$ are
the eigenfunctions of the operator 
$\vec\alpha\cdot \left(\vec k-g\vec A^a{\tau^a\over2}\right)$ 
with eigenvalues
$E_A=\alpha\left(\sqrt{{\vec k}^2+h^2}+\beta h\right)$ 
each of which is doubly degenerate.
($A$ denotes collectively $\alpha,\beta=\pm$). 
After Fourier-transforming in $t$, we find that the kinetic 
term in the action is
\begin{equation}
S_{0}=\sum_{A,i}\int
{d\omega\over2\pi}{d^2\vec k\over(2\pi)^2}
{\tilde \psi}^{\dagger}_{Ai}(\omega,k)
(\omega-E_A){\tilde \psi}_{Ai}(\omega),
\qquad i=1,2,
\end{equation}
where $\tilde\psi_{Ai}(\omega,k)$ is the Fourier transform of 
$\psi_{Ai}(t,k)$ and the spinors $u_{Ai}$ are chosen to be 
normalized as
$u^\dagger_{Ai}u_{Aj}= \delta_{AB}\delta_{ij}$.

When $|\vec k|\ll h$, the effect of the external field is important.
It changes the energy spectrum drastically. To derive the low energy
effective action, we integrate out the modes with 
$|\omega|,|\vec k|>h$. Then, the kinetic term in the effective 
action becomes
\begin{equation}
S_{0}=\sum_{\alpha,i}\int_{|\omega|, |\vec k|<h}
{d\omega\over2\pi}{d^2\vec k\over(2\pi)^2}
\left[{\tilde \psi}^{\dagger}_{\alpha i,-}(\omega-\alpha{{\vec
k}^2\over 2h}){\tilde \psi}_{\alpha i,-}
+{\tilde \psi}^{\dagger}_{\alpha i,+}(\omega-
2\alpha h){\tilde \psi}_{\alpha i,+}\right]+\cdots,
\end{equation}
where we have expanded the energy eigenvalues $E_A$ in powers of 
momentum and the elipsis denotes the terms with higher powers of 
momenta.

To determine the scaling dimensions of the fermion modes, 
consider a transformation,
\begin{equation}
\omega\to s\omega,\quad \vec k\to s^{1/2} 
\vec k\quad {\rm with} \quad s<1.
\end{equation}
Since the kinetic term must be invariant under the scaling
transformation, we find ${\tilde \psi}_{\alpha i,-}$ has scaling
dimension $-3/2$, while ${\tilde \psi}_{\alpha i,+}$ has $-1$.
Namely, in the coordinate space, $\psi_{\alpha i,\pm}(\vec r,t)$,
the Fourier transform of ${\tilde \psi}_{\alpha i,\pm}$,
has scaling dimension $1/2$ and $1$, respectively.
We see that the component $\psi_{\alpha i,-}$ is more relevant 
in the infrared region than the component $\psi_{\alpha i,+}$.
Because of the change in the energy spectrum due to the
external field, the scaling dimension of fermion field changes for
the component $\psi_{\alpha i,-}$ at low energy as if the 
spacetime is dimensionally reduced from $2+1$ to $1+1$.
Because of this dimensional reduction, the
four-Fermi interaction of $\psi_{\alpha i,-}$ is marginal.
Now, let us further integrate out the modes of $sh<\omega<h$,
$s^{1/2}h<|\vec k|<h$ to find the change in the four-Fermi 
interaction. The one-loop correction to the four-Fermi coupling 
in the leading order  in $1/N$ is given then 
\begin{eqnarray}
\delta G= {-iG^2\over N}\int_{sh<\omega<h, s^{1/2}h<|\vec k|<h}  
{\rm Tr}\left[ S(k)  S(k)\right]=-{G^2 h\over \pi }\ln s.
\end{eqnarray}
In terms of the dimensionless coupling $\lambda$,  we find
\begin{equation}
\beta(\lambda)=s{\partial \lambda\over \partial s}
=-{1\over \pi}{\lambda}^2.
\end{equation}
Therefore, the four-Fermi interaction becomes a relevant operator 
at quantum level if $G>0$, which at low energy leads to strong 
attraction between fermion and antifermion to form a condensate. 
The dynamical fermion mass, the RG invariant scale, is
$m_{\rm dyn}\simeq h\exp[-\pi/ hG)]$, which is, in  leading
order in the coupling expansion, the same as the dynamical mass we
obtained by solving the gap equation~(\ref{rg}) for weak coupling
$\lambda$ at $\mu=h$, where the RG analysis is reliable.

The (2+1)-dimensional $SU(2)$ gauge theory is 
more complicated than the 3D NJL model, since
not only the fermions but also the gauge field propagate
nontrivially under the external uniform chromomagnetic field.
The gauge field propagator (in Feynman gauge) in momentum 
space is determined as a solution to the equation:
\begin{equation}
\left[{{\cal P}}_{\mu}{{\cal P}}^{\mu} g_{\alpha\beta}
-2igF^a_{\alpha\beta}I^a\right]
S^{\beta\gamma}(p)=-i\delta_\alpha^\gamma.
\end{equation}
where ${{\cal P}}_{\mu}=p_{\mu}-gI^{a}A^{a}_{\mu}$ and $I^a$ 
are the generators in the adjoint representation.
As is seen, to solve this equation one has to invert a matrix
with double indices, Lorentz and internal. The structure in
Lorentz indices, however, can be simplified considerably. That is 
due to the fact that the only nondiagonal Lorentz matrix appearing 
in the equation is $F^3_{\alpha\beta}$, and it commutes with
$g_{\alpha\beta}$. So, we represent the gluon propagator through the
projectors which diagonalize the $F^3_{\alpha\beta}$-matrix:
\begin{eqnarray}
S^{\beta\gamma}(p) = -i\sum_{j=-1}^{1}S^{j}(p) A^{\beta\gamma}_{j},
\end{eqnarray}
where
\begin{eqnarray}
A^{\beta\gamma}_{0}=g^{\beta\gamma} + \frac{F^{3\beta\nu}
F_{~\nu}^{3~~\gamma}}{H^2},
&&\quad
A^{\beta\gamma}_{\pm 1}=\frac{1}{2}\left[
\mp i\frac{F^{3\beta\gamma}}{H}
-\frac{F^{3\beta\nu}F_{~\nu}^{3~~\gamma}}{H^2}\right],\\
\sum_{j=-1}^{1} A^{\beta\gamma}_{j} = g^{\beta\gamma},
&&\quad
F^{3\beta}_{~~\nu}A^{\nu\gamma}_{j}= i H j A^{\beta\gamma}_{j}.
\end{eqnarray}
Thus, we obtain the equation for $S^j$
\begin{eqnarray}
\left[p^2-4h^2\left((I^1)^2+(I^2)^2\right)
-4h(\vec{p}\cdot\vec{I}) + 8jh^2I^3\right] S^{j} = 1,
\end{eqnarray}
which is easily solved by a simple inversion of $3\times 3$ 
matrices. The result reads:   
\begin{eqnarray}
S^{j} = \frac{a_j +b(\vec{p}\vec{I}) +c(I^3)^2 + jdI^3
+f\left[(\vec{p}\vec{I}) -2hjI^3\right]^2}{E_j},
\end{eqnarray}
where
\begin{eqnarray}
&&a_j=(p^2-4h^2)^2-16h^2{\vec p}^2-64h^4j^2,\quad
b=4h(p^2-4h^2),\\
&&c=-4h^2(p^2-4h^2),\quad
d=-8h^2(p^2-8h^2),\quad
f=16h^2,\\
&&E_j=(p^2-8h^2)(p^2-4h^2)^2-16 h^2{\vec p}^2(p^2-4h^2)
-64h^4j^2(p^2-8h^2).
\end{eqnarray}

We find that under the external chromomagnetic field the gauge 
field is not massless, instead its propagator has poles at 
nonzero $p$, all of order $h$, satisfying
$(p^2-8h^2)(p^2-4h^2)^2-16h^2\vec{p}^2(p^2-4h^2)-64h^4j^2(p^2-8h^2)
=0,\quad j=0,\pm 1$. Seven of these poles correspond to massive
modes while the remaining two describe tachyonic modes signaling
instability of the perturbative vacuum in the external chromomagnetic 
field (\ref{potential}) (analog of the Nielsen-Olesen instability in
four-dimensional Yang-Mills theory for abelian configurations of
chromomagnetic field~\cite{Nielsen}).\footnote{For a thorough 
investigation of the instability of non-Abelian fields given by
constant potentials see \cite{Leutwyler}.} Since we are interested 
in infrared dynamics of fermions with energy and momenta 
$E,|\vec p|\ll h$, the presence of tachyonic modes of order $h$ is 
not very important. In fact we will be working in an effective low 
energy theory where $h$ serves as ultraviolet cutoff. Further we 
need the gauge propagator which is approximately constant for 
$E, |\vec p|\sim 0$,
\begin{eqnarray}
S^{ab}_{\alpha\beta}&=&\frac{i}{8h^2}g_{\alpha\beta}
\left(2\delta^{ab}-\delta^{3a}\delta^{3b} \right)
+\frac{i}{6h^2}\frac{F_{\alpha\beta}}{H}\varepsilon^{3ab}
+\frac{i}{3h^2}\frac{(F^2)_{\alpha\beta}}{H^2}
\left(\delta^{ab}-\delta^{3a}\delta^{3b}\right).
\label{glu0}
\end{eqnarray}
 Since the gauge field is massive, the gauge field exchange
interaction will generate a four-Fermi interaction
at low energy $E\ll h$:
\begin{equation}
O_4(x)=-{g^2\over 48h^2}\left[
\left(\bar\psi\psi\right)^2+\left(\bar\psi i\gamma_5\psi\right)^2
+\left(\bar\psi \gamma_3\psi\right)^2
\right]+\cdots,
\label{fourfermi}
\end{equation}
where the ellipsis denotes other than NJL interaction terms. For a 
while we assume that only NJL-like interactions play important role 
in the generation of a dynamical mass. Then, the effective four-Fermi 
interaction generated by the massive gauge field is repulsive 
($-{g^2\over 48h^2}<0$). Therefore, even though the four-Fermi 
interaction for the $\tilde\psi_{\alpha i,-}$ component of the 
fermion is marginal as we have shown in RG analysis of the 3D NJL 
model, its $\beta$-function is positive and it becomes an irrelevant
operator at one-loop. Thus we may conclude that the
chromomagnetic field does not induce chiral symmetry breaking
in 3D QCD.

Since, while deriving the effective four-fermion interaction, we 
omitted terms other than those in (\ref{fourfermi}) in order to 
apply the RG analysis, below we confirm our conclusion by analysis of 
the gap equation for a fermion dynamical mass. There are two 
equivalent approaches one can follow in studying mass generation, 
that of  Schwinger-Dyson equations~\cite{leung,my}, and that of 
Bethe-Salpeter (BS) equation~\cite{gusynin}. We consider here the 
BS equation for a gapless Nambu-Goldstone (NG) boson, composed 
of a fermion and antifermion, which must appear when the chiral 
symmetry is spontaneously broken. 

The homogeneous BS equation for the NG bound state reads:
\begin{eqnarray}
\rho=-4\pi\alpha \int\frac{d^3p}{(2\pi)^3}
{\tau^a\over2}\gamma^\mu S(p)\rho S(p){\tau^b\over2}\gamma^\nu 
S^{ab}_{\mu\nu},
\label{BSeq}
\end{eqnarray}
where $\alpha=g^2/(4\pi)$, and
the Bethe-Salpeter wave function for the bound state is
given in general as
\begin{eqnarray}
\rho=\gamma^5\left(A+\tau^3\sigma^{12}B
+(\vec{\gamma}\cdot\vec{\tau})C \right).
\label{rho}
\end{eqnarray}
Since the dynamics responsible for fermion mass generation
is due to momenta near $p=0$,
we can neglect the momentum dependence in the gauge field
propagator in the BS equation. Thus, the functions $A,B,C$
become momentum independent and Eq.(\ref{BSeq}) turns out to
be an algebraic system of equations.

By direct substitution of (\ref{rho}) in the BS equation, we
observe that the equation for coefficient $C$ decouples from the
set of equations for $A$ and $B$. The latter is:
\begin{eqnarray}
A&=&\frac{i\pi\alpha}{24h^2}
\int\frac{d^3p}{(2\pi)^3}
\frac{\left[6h^2+13(p^2-m^2)\right]A+\left[6h^2
-16(p^2+2\vec{p}^2-m^2) \right]B}{Q_1(p)Q_2(p)}, \\
B&=&\frac{i\pi\alpha}{24h^2} \int\frac{d^3p}{(2\pi)^3}
\frac{\left[78h^2-16(p^2-m^2)\right]A+\left[78h^2
-23(p^2+2\vec{p}^2-m^2)\right]B}{Q_1(p)Q_2(p)}.
\end{eqnarray}
In the limit $m\to 0$, this reduces to the following
\begin{eqnarray}
A&\simeq&-\frac{\alpha}{64h} \ln\left(\frac{h}{m}\right)(A+B),\\
B&\simeq&-\frac{13\alpha}{64h} \ln\left(\frac{h}{m}\right)(A+B),
\end{eqnarray}
or in other words, we obtain
\begin{eqnarray}
1&\simeq&-\frac{7\alpha}{32h} \ln\left(\frac{h}{m}\right).
\end{eqnarray}
Thus, there is no solution satisfying the condition $m\ll h$,
which agrees with the result of RG analysis (analysis of an 
equation for $C$-coefficient shows no nontrivial solution, too).

While  3D QCD itself is known to be unstable 
in perturbation theory~\cite{cornwall},
and might lead to generating a mass for Yang-Mills 
fields nonperturbatively~\cite{nair}, the presence of an external 
chromomagnetic field enhances this instability. A chromomagnetic field 
influences the behavior of charged vector particles more strongly 
than that of fermions and further study is needed to 
make decisive conclusion about the magnetic catalysis in 3D QCD. 
Perhaps, simultaneous study of generating chiral condensate and 
a condensate of vector particles would shed light on this problem. 
The absence of chiral condensate in a chromomagnetic field in 3D QCD 
may cast doubts upon straightforward application of the results of 
a magnetic catalysis in QED4~\cite{gusynin,leung,my} 
to QCD4~\cite{smilga}. 

In conclusion, we investigated the effect of a constant external
non-Abelian field on chiral symmetry breaking in a (2+1)-dimensional
Nambu-Jona-Lasinio model and in 3D QCD. We studied the effect
by gap equation analysis and RG analysis and found that in the
(2+1)-dimensional NJL model chiral symmetry breaking occurs
for any weak coupling constant 
(in accordance with earlier findings~\cite{nonabnjl})
but catalysis of chiral symmetry breaking does not
occur in 3D QCD.

\acknowledgments

We are grateful to V.A. Miransky for helpful remarks and F. Hawes 
for correcting the manuscript. The final version of the paper was 
written when one of us (V.P.G.) was staying at the Special Research 
Centre for Subatomic Structure of Matter, the University of Adelaide, 
Australia. He would like to thank Prof. A. Thomas and Dr. A. Williams 
and all other members of the Centre for hospitality. This work 
is  
supported in part by Swiss National Science Foundation grant 
CEEC/NIS/96-98/7 IP 051219 (V.P.G.), 
by
the academic research
fund of Ministry of Education, Republic of Korea, Project
No. BSRI-97-2413 (D.K.H.), by the KOSEF through SRC program of
SNU-CTP (D.K.H.), by the U.S.
Department of Energy grant \#DE-FG02-84ER40153 (I.A.Sh.), 
 and by Foundation of Fundamental 
Researches of Ministry of Sciences of the Ukraine under grant 
N 2.5.1/003 (V.P.G., I.A.Sh.). 
One of the authors (D.K.H.) wishes to acknowledges the partial financial 
support of the Korea Research Foundation made in the program 
year of 1997.


\begin{references}

\bibitem{qcd}For recent review, see C. D. Roberts and A. G. Williams,
Progr. Part. Nucl. Phys. {\bf33}, 477 (1994).

\bibitem{sd}
V. A. Miransky, {\it Dynamical Symmetry Breaking in Quantum Field
Theories} (World Scientific, 1993), and references therein.

\bibitem{thooft}S. Mandelstam, Phys. Reports {\bf 23}, 245 (1976);
G. 't Hooft, in {\it 1981 Carg\'ese Summer School Lectures on
Fundamental Interactions, NATO Adv. Study Inst. Series B: Phys.},
vol. 85, edited by M. L\'evy {\it et al.}
(Plenum Press, New York, 1982); Nucl. Phys. {\bf B190}, 455 (1981).

\bibitem{seiberg}N. Seiberg and E. Witten, Nucl. Phys. B{\bf426}, 
19 (1994); B{\bf 431}, 484 (1994).

\bibitem{banks}T. Banks and A. Casher, Nucl. Phys. {\bf B169},
103 (1980); E. Marinari. G. Parisi, and C. Rebbi, Phys. Rev.
Lett. {\bf 47}, 1795 (1981); H. Leutwyler and A. Smilga, D{\bf46},
5607 (1992).

\bibitem{NJL}V. P. Gusynin, V. A. Miransky, and I. A. Shovkovy,
Phys. Rev. Lett. {\bf73}, 3499 (1994); Phys. Rev. D{\bf52}, 4718
(1995); Phys.Lett. B{\bf349}, 477 (1995).

\bibitem{gusynin}V. P. Gusynin, V. A. Miransky, and I. A. Shovkovy,
Phys. Rev. D {\bf 52}, 4747 (1995); Nucl. Phys. B {\bf 462}, 249
(1996); V.P. Gusynin and I.A. Shovkovy, Phys. Rev. D {\bf 56}, 5251 
(1997).\\
 V. P. Gusynin, ``Magnetic catalysis of chiral symmetry 
breaking in QED at finite temperature'', hep-ph/9709339.

\bibitem{leung}C. N. Leung, Y. J. Ng and A. W. Ackley,
 Phys. Rev. D {\bf 54}, 4181 (1996); D.-S. Lee, C. N. Leung,
Y. J. Ng, Phys. Rev. D {\bf 55}, 6504 (1997).

\bibitem{my}D. K. Hong, Y. Kim, and S. Sin, Phys. Rev. D {\bf 54},
7879 (1996); D. K. Hong, Seoul National University Report,
SNUTP 97-053, hep-ph/9707432.

\bibitem{KKK}S. Kawati, G. Konisi, and H. Miyata, Phys. Rev. 
D{\bf28}, 1537 (1983); 
S.P. Klevansky and R.H. Lemmer, Phys. Rev., D{\bf39}, 3478 (1989);
I. V. Krive and S. A. Naftulin, Phys. Rev. D{\bf46}, 2737 (1992).

\bibitem{Klim}K. G. Klimenko, Z. Phys. C{\bf54}, 323 (1992).

\bibitem{mavromatos}K. Farakas and N. E. Mavromatos, 
``Gauge-theory approach to planar doped antiferromagnets and 
external magnetic fields'', cond-mat/9710188.

\bibitem{witten}E. Witten, Nucl. Phys. B{\bf145}, 110 (1978).

\bibitem{Brown} L. S. Brown and W. I. Weisberger, Nucl. Phys.
B{\bf157}, 285 (1979).

\bibitem{Leutwyler}H. Leutwyler, Nucl. Phys. B{\bf179}, 129 (1981).

\bibitem{Ebert}D. Ebert and V. Ch. Zhukovsky,``Chiral phase
transitions in strong chromomagnetic fields at finite temperature 
and dimensional reduction", hep-ph/9701323.

\bibitem{nonabnjl}K. G. Klimenko, B. V. Magnitsky, 
and A. S. Vshivtsev, Nuovo Cim. A{\bf107}, 439 (1994); 
Teor. Mat. Fiz. {\bf101}, 391 (1994).

\bibitem{Nielsen}H. Nielsen and P. Olesen, Nucl. Phys. B{\bf144},
376 (1978).

\bibitem{cornwall}J. M. Cornwall, ``On one-loop gap equations for 
the magnetic mass in $d=3$ gauge theory'', hep-th/9710128.

\bibitem{nair}D. Karabali and V. P. Nair, Int. J. Mod. Phys. 
A{\bf12}, 1161 (1997).

\bibitem{smilga}I. A. Shushpanov and A. V. Smilga, 
Phys. Lett. B{\bf 402}, 351 (1997).

\end{references}
\end{document}